\documentclass{article}
\usepackage[utf8]{inputenc}
\usepackage[normalem]{ulem}
\usepackage{graphicx} 
\usepackage{authblk}
\usepackage{xcolor}
\usepackage{tablefootnote}
\usepackage{cite}
\usepackage{amsmath,amssymb,amsfonts}
\usepackage{mathtools}
\usepackage{algorithmic}
\usepackage{textcomp}
\usepackage{relsize}

\usepackage[font=small,labelfont=bf]{caption}
\usepackage[square,numbers]{natbib}
\usepackage{hyperref}
\usepackage{booktabs}

\newcounter{bla}

\newcommand{\be}{\begin{equation}}

\newcommand{\ee}{\end{equation}}
\definecolor{codegreen}{rgb}{0,0.6,0}
\definecolor{codegray}{rgb}{0.5,0.5,0.5}
\definecolor{codepurple}{rgb}{0.58,0,0.82}
\definecolor{backcolour}{rgb}{0.95,0.95,0.92}

\usepackage{listings}
\definecolor{mGreen}{rgb}{0,0.6,0}
\definecolor{mGray}{rgb}{0.5,0.5,0.5}
\definecolor{mPurple}{rgb}{0.58,0,0.82}
\definecolor{backgroundColour}{rgb}{0.95,0.95,0.92}
\lstdefinestyle{mystyle}{
    backgroundcolor=\color{backgroundColour},   
    commentstyle=\color{mGreen},                
    keywordstyle=\color{mPurple}\bfseries,      
    numberstyle=\tiny\color{mGray},             
    stringstyle=\color{red},                    
    basicstyle=\ttfamily\footnotesize,
    breaklines=true,
    captionpos=b,
    keepspaces=true,
    numbers=left,
    numbersep=5pt,
    showspaces=false,
    showstringspaces=false,
    showtabs=false,
    tabsize=2,
    frame=single,
    framerule=0.5pt,
    rulecolor=\color{mGray}                     
}

\title{Hybrid Fourier Neural Operator–Plasma Fluid Model for Fast and Accurate Multiscale Simulations of High Power Microwave Breakdown}

\author[1,2]{Kalp Pandya}
\author[1]{Pratik Ghosh}
\author[1]{Ajeya Mandikal}
\author[1]{Shivam Gandha}
\author[1,2]{Bhaskar Chaudhury\thanks{Corresponding author: \texttt{bhaskar\_chaudhury@dau.ac.in}}}

\affil[1]{Group in Computational Science and HPC, DA-IICT, Dhirubhai Ambani University, Gandhinagar, India, 382007}
\affil[2]{Smart Energy Learning Center, DAU, Gandhinagar, India, 382007}

\date{ }

\begin{document}
\maketitle

\begin{abstract}
Modeling and simulation of High Power Microwave (HPM) breakdown, a multiscale phenomenon, is computationally expensive and requires solving Maxwell’s equations (EM solver) coupled with a plasma continuity equation (plasma solver).
In this work, we present a hybrid modeling approach that combines the accuracy of a differential equation-based plasma fluid solver with the computational efficiency of FNO (Fourier Neural Operator) based EM solver. Trained on data from an in-house FDTD-based plasma-fluid solver, the FNO replaces
computationally expensive EM field updates, while the plasma solver governs the dynamic plasma response. The hybrid model is validated on microwave streamer formation, due to diffusion-ionization mechanism, in a 2D scenario for unseen incident electric fields corresponding to entirely new plasma streamer
simulations not included in model training, showing excellent agreement with FDTD-based fluid simulations in terms of streamer shape, velocity, and temporal evolution. This hybrid FNO based strategy delivers significant acceleration of the order of 60X compared to
traditional simulations for the specified problem size and offers an efficient alternative for computationally demanding multiscale and multiphysics simulations involved in HPM breakdown.  Our work also demonstrate how such hybrid pipelines can be used to seamlessly to integrate existing C-based simulation codes with Python-based machine learning frameworks for simulations of plasma sceince and engineering problems.
\end{abstract}

\vspace{2pc}

\section{Introduction}\label{sec: Introduction}
High-power microwave (HPM) breakdown in gases is an inherently complex and nonlinear phenomenon, wherein intense electromagnetic (EM) waves ionize a gas, initiating plasma formation and evolution \cite{ADMacdonald1966, Takahashi2024}. HPM breakdown has been investigated both computationally and experimentally, for the last several decades, driven by a wide range of applications, including plasma propulsion, ignition in fusion devices, communications
technology, deep-space communications, aerospace applications, flow control, electromagnetic warfare and electromagnetic shielding among others ~\cite{Nusinovich1997, takahashiAIP, Oda2020, Fukunari2019,Khodataev2008,Saifutdinov2019, BChaudhuryPRL2010, olivier2022, JBenfordHPMdeepspapp2008,HPMeffectUWBtransmission2010, guolin2010, scirep23,Hidaka2009,Takahashi-2019,MFukunari2018,PratikIMaRC2021}. 
The types of gas discharge observed experimentally include streamer, overcritical, subcritical, volumetric, and initiator-attached forms ~\cite{Khodataev2008,temkin,Aleksandrov2006,PBulat2021, Saifutdinov2021, Hidaka2008,JBMichael-2010,tabata24,suzuki2025,MTakahashietal2017}. 
In recent years, there has been growing interest in HPM breakdown experiments under high-frequency (100–300 GHz) and high-pressure (10–760 torr) conditions, where the resulting filamentary plasmas can be harnessed for efficient energy coupling, ignition enhancement, and flow manipulation~\cite{Schaub2016,Aleksandrov2006,Oda2020,Yoda2006,AMCook2011,PBulat2021}.
HPM-induced plasmas at high pressures and high frequencies are increasingly seen as enabling technologies for next-generation energy-efficient propulsion technologies and green energy applications including aerodynamic flow control, plasma-assisted combustion, energy deposition in supersonic and hypersonic gas-dynamic flows, beamed-energy propulsion and flame stabilization~\cite{vedenin,fukunaribook,PBulat2021,raja2021}. 

Modeling and simulation of HPM breakdown are crucial for understanding the fundamental physics of EM wave–plasma interactions, optimizing experimental configurations, and developing predictive tools for engineering applications. Several semi-analytical models and computational approaches have been employed to study the spatial and temporal dynamics of plasma formation during HPM breakdown, offering deeper insights into key processes such as nonlinear ionization, plasma evolution, and energy exchange among the wave, plasma, and gas ~\cite{BChaudhuryPRL2010, Fukunari2019,NamVerboncoeur2009,BChaudhury2010,Nakamura2018,Saifutdinov2021,Saifutdinov23,BCHAUDHURY2018,Suyan-2016,Konstantinos-2015,Kourtzanidis20143DHPM,boeuf2017, vedenin,BChauduryIEEEplasma2010,Ref1SYan2018,Ref3Zhao2011,Semenov2015,HAMIAZ2020,QZHAO2011}.\\
At high pressures, where plasmas are highly collisional and the mean free path is much smaller than the plasma dimensions, the plasma can be effectively treated as a continuum. In such cases, traditional computational approaches often employ fluid models that couple Maxwell’s equations with plasma continuity equations~\cite{BChaudhuryPRL2010,BChaudhury2011}. Particularly in high-frequency regimes, the finite-difference time-domain (FDTD) method is widely used to solve Maxwell's equations due to its simplicity and second-order accuracy. The coupled FDTD–fluid solver framework has successfully reproduced key experimental observations, including microwave streamer evolution, plasma front propagation, filamentation, and the formation of self-organized plasma structures in high-pressure environments~\cite{BChaudhury2010}.
A common characteristic of HPM breakdown at atmospheric pressure is the formation of an initial plasmoid around one or more seed electrons, which elongates in the direction parallel to the incident field and evolves into a plasma filament, often referred to as a “microwave streamer"~\cite{BChaudhury2011}. Microwave breakdown studies become more tractable when the field above breakdown is confined to a small volume, enabling controlled observation of a single streamer elongated along the direction of the incident field. Such configurations have been used in experiments and also serve as established benchmarks for investigating breakdown physics and validating computational models~\cite{boeuf2017,vedenin,PBulat2021}.\\
HPM breakdown is inherently a multiscale, multiphysics problem due to the presence of different time and space scales. Especially at high frequencies, above 100 GHz, it involves phenomena ranging from sub-wavelength EM field oscillations occurring on picoseconds timescales to plasma density evolution that spans over tens of nanoseconds to microseconds. Similarly, millimeter- to micrometer-scale density and electric-field gradients exist. Accurately capturing such a wide range of spatial and temporal scales demands very fine grids to resolve sharp density and electric field gradients, and very small time steps to satisfy the Courant–Friedrichs–Lewy (CFL) condition. This leads to enormous computational costs. To address the limitations of traditional computational approaches, various strategies have been explored, including advanced parallelization techniques and adaptive or selective mesh refinement, aimed at reducing simulation time without compromising accuracy~\cite{BCHAUDHURY2018,Pratik2020}. 
Despite the use of these techniques, simulating realistic large-scale two or three-dimensional scenarios over extended time scales (microseconds) remains nearly intractable with conventional methods. This underscores the need for alternative strategies that can more efficiently meet the substantial computational demands of realistic HPM breakdown simulations with limited computational resources. \\
Deep learning is increasingly being used to accelerate the simulation of physics and engineering problems, particularly for dynamical systems that evolve in time, where conventional numerical solvers for PDEs and ODEs can be prohibitively expensive~\cite{dlbook,review-rmp,dlreview,dlfluid}. Approaches such as physics-informed neural networks (PINNs) include governing equations into the loss function to learn time-dependent solutions directly, while neural operators like the Fourier Neural Operator (FNO) learn mappings between function spaces~\cite{pinn,pinnreview}. These methods can serve as surrogates for traditional solvers, offering orders-of-magnitude speedups for parametric studies, optimization tasks and real-time investigations. In plasma physics and electromagnetics, deep learning has been applied to mimic full-wave solvers for microwave–plasma interactions and to reconstruct plasma density from scattered EM signals with high accuracy~\cite{pratiktmm,pratikjpd,tap2025}. Recent studies focused on applications of machine learning in the simulations of low temperature plasmas and fusion plasmas demonstrates the potential of deep learning and data-driven approaches in these areas~\cite{2022review,fno-fusion}. The integration of machine learning and physics-based modeling is emerging as a promising approach toward efficient and scalable simulation of complex time-evolving systems.
In this work, we propose a hybrid pipeline, involving integration of a differential equation (DE) based numerical approach with a data-driven deep learning (DL) model to improve
computational efficiency of plasma fluid modeling. As a proof of concept, we apply this hybrid approach using FNO (Fourier neural operator) to simulate HPM breakdown in air in a 2D computational setup, capturing the spatiotemporal evolution of plasma and the formation of plasma streamers~\cite{BChaudhury2011}. A standing wave is produced by two identical, linearly polarized waves incident from opposite sides of a 2D rectangular domain, with the streamer initiated by seed electrons placed at the field maximum. The simulations provide the spatio-temporal evolution of plasma density and electromagnetic fields. In this study, the simulations are limited to time durations of a few hundred nanoseconds, which are shorter than the characteristic time scale of gas heating. Accordingly, the gas temperature is assumed to remain constant, as reported in the literature. The traditional fluid approach applied for this problem involves coupling of an electromagnetic (EM) solver and a plasma solver. The EM solver, based on Maxwell’s
equations, models the interaction between microwaves and plasma, while the plasma solver addresses the plasma continuity equation. Notably, the EM solver consumes approximately more than $99\%$
of the overall simulation time, whereas the plasma solver accounts for less than $1\%$, underscoring
the potential for optimizing the EM solver via deep learning to significantly accelerate such simulations. 
The objective of this paper is to assess the efficiency and accuracy of hybrid computational approaches in comparison with conventional differential equation based methods, and to demonstrate their application through a case study of microwave streamer formation during HPM breakdown. The results provide an estimate of the computational speedup achievable for a representative 2D simulation, noting that the speedup may vary from problem to problem.\\
The overall structure of this paper is organized as follows. Section~\ref{sec: HPM Breakdown Simulation} provides an in-depth description of the physical and widely used computational model of HPM breakdown. Subsection~\ref{subsec: physical model HPM} outlines the physical model, while subsection~\ref{subsec: numerical fdtd simulation} presents its discretized form along with the iterative coupling of EM and plasma density solvers. The proposed FNO-based hybrid model is detailed in section~\ref{sec:hybrid_fno_framework}, with implementation aspects of invoking C language-based solvers via a Python controller discussed in subsection~\ref{subsec: py_c_interfacing}. Section~\ref{sec: FNO model preparation} describes the creation and training of the FNO model through three subsections - subsection~\ref{subsec: data_generation} covers the preparation of training, validation, and testing datasets; subsection~\ref{subsec: model architecture} explains the model architecture; and subsection~\ref{subsec: training loop} outlines the training loop for a single batch. Section~\ref{sec: results_validation} presents the results and validation, with subsection~\ref{subsec:fno_performance} evaluating the performance of FNO model in substituting conventional EM solver, and subsection~\ref{subsec:hybrid_performance} comparing the simulation performance of the FDTD plasma fluid model and the FNO-based hybrid model. Finally, section~\ref{sec: Conclusion} summarizes the key findings and contributions of this work.

\section{Physical and Conventional Computational Model of HPM Breakdown}
\label{sec: HPM Breakdown Simulation}

\subsection{Physical Model of HPM Breakdown}
\label{subsec: physical model HPM}
\begin{figure}[htbp]
    \centering
    \includegraphics[width=1\textwidth]{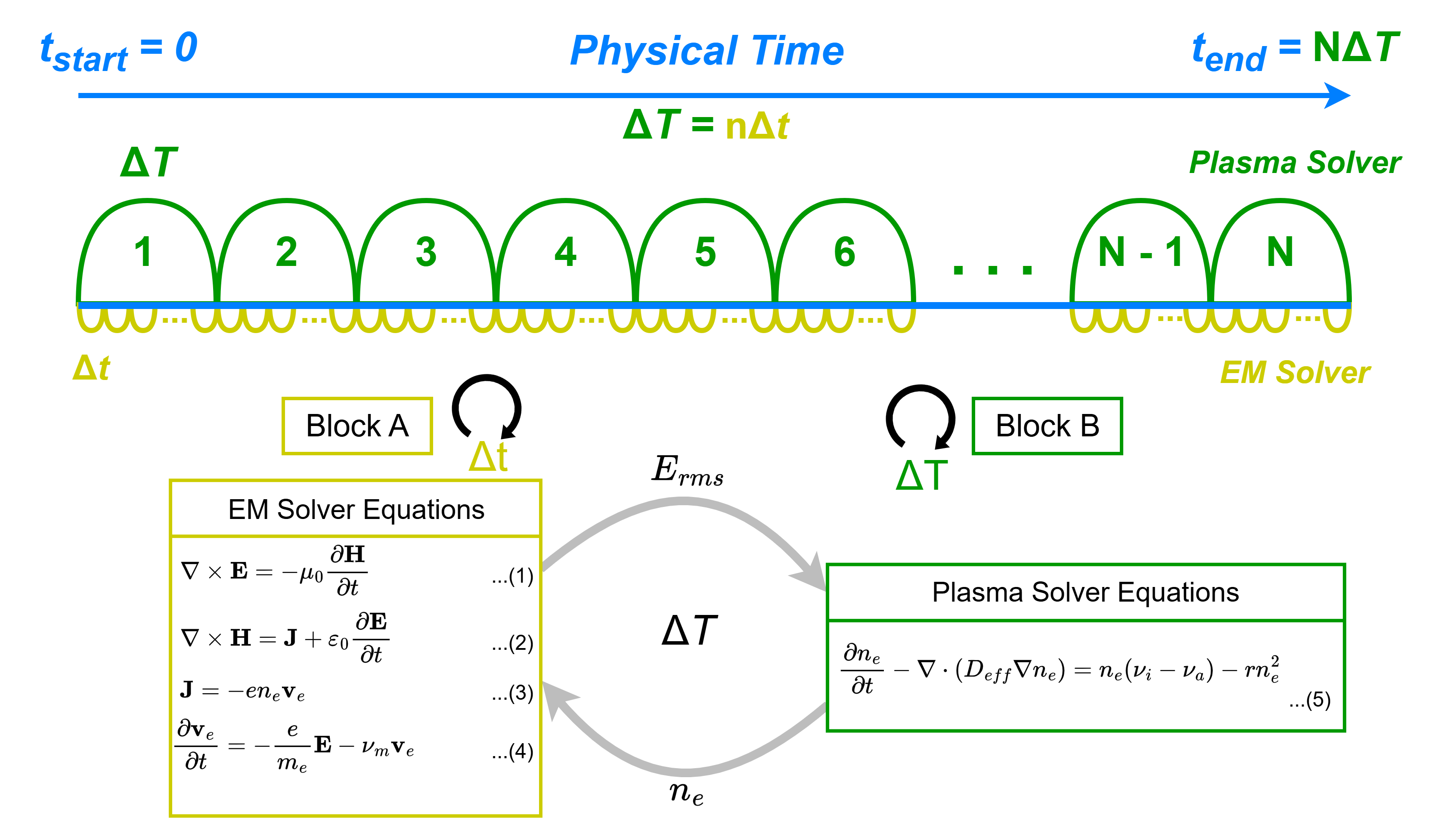}
    \caption{The figure illustrates the FDTD-based plasma fluid model showing the iterative coupling between the EM and plasma solvers over the full physical time period ($t_{end}-t_{start}$), with each operating at different time steps. The plasma solver is invoked after "n" iterations of the EM solver, where "n" corresponds to the number of EM solver iterations in one EM wave period.}
    \label{fig:HPM_Breakdown}
\end{figure}

Two primary physical phenomena govern HPM breakdown, the interaction of EM wave with plasma described by Maxwell's equations (Eqs. 1,2 in Fig.\ref{fig:HPM_Breakdown}; Block A- EM Solver) and plasma dynamics governed by the plasma continuity equation (Eq. 5 in Fig. \ref{fig:HPM_Breakdown}; Block B - Plasma Solver). 
The physical model of HPM breakdown described here follows the models presented in \cite{BChaudhuryPRL2010,BChaudhury2010,BChaudhury2011}, wherein the coupling between Maxwell's equations and plasma continuity equation occurs through the electron current density ($\mathbf{J}$).
Figure \ref{fig:HPM_Breakdown} illustrates the iterative coupling between the EM and plasma solvers over the entire simulation. In EM solver (Block A  in Fig. \ref{fig:HPM_Breakdown}), $\mathbf{E}$, $\mathbf{H}$, $\mu_0$ and $\varepsilon_0$ represent the electric and magnetic fields of the EM wave, magnetic permeability and electric permittivity of free space respectively.
The plasma is assumed to be quasi-neutral, and the ion contribution to the total current density is negligible and, therefore, neglected. The electron (plasma) current density $\mathbf{J}$ is described by equation (3) in figure \ref{fig:HPM_Breakdown}. 
Here, $e$ denotes the electron charge, $n_e$ represents the electron density (in $m^{-3}$), and $\mathbf{v}_e$ indicates the mean electron velocity computed using equation (4) in figure \ref{fig:HPM_Breakdown}, where $m_e$ is the electron mass, and $\nu_m$ is the electron-neutral collision frequency. For air plasma considered in this study, the collision frequency is expressed as $\nu_m = 5.3 \times 10^{9} \times p$, where $p$ is the ambient pressure in torr \cite{BChaudhury2011}.

The electron continuity equation in Plasma solver (figure \ref{fig:HPM_Breakdown}) is averaged over one EM wave cycle, resulting in a flux divergence term that includes only diffusion, with no convective contribution. This simplification is justified by the significant separation of time scales, as the plasma density evolves much more slowly compared to the period of the EM wave. In other words, in the high-frequency regime, plasma propagation is dominated by diffusion rather than drift, therfore, only the diffusive term is retained in the flux divergence term of the continuity equation ~\cite{BChaudhury2011}. This equation incorporates essential transport phenomena, including diffusion $D_{eff}$, ionization, attachment and recombination $r$ processes. The effective diffusion coefficient ($D_{eff}$) is described extensively in \cite{BChaudhury2011}. In this study, the electron temperature is assumed to be approximately constant at 2 eV, as described in ~\cite{BChaudhury2011,BChauduryIEEEplasma2010}.
The ionization frequency ($\nu_i$) and 
the attachment frequency ($\nu_a$) are computed from solutions of the Boltzmann equation under a uniform DC reduced electric field $E_{eff}/p$, which is a standard approximation for microwave breakdown at high pressure, where $p$ 
denotes the pressure in Torr. 
The nonlinear interaction between the EM wave and plasma density primarily occurs through this ionization frequency. These parameters, $D_{eff}$, $\nu_i$, $\nu_a$ and electron–ion recombination coefficient $r$, form critical transport coefficients governing plasma dynamics. The computational model is based on the classical effective field approximation, which assumes that electron transport properties depend solely on the local effective field $E_{effective}$ given by:
\begin{equation}
E_{eff}=\sqrt{\frac{E_{rms}^{2}}{(1+\omega^{2}/\nu_{m}^{2})}}
\label{eq:eff}
\end{equation}
Here $E_{rms}$ is the total rms field at the considered location, and $\omega$ is the angular frequency of the incident EM wave. More details of the EM–plasma fluid model are available in the published literature~\cite{BChaudhury2011,BChauduryIEEEplasma2010,BCHAUDHURY2018,ghosh2023efficient}.

    \subsection{Numerical Implementation of the Fluid Model for HPM Breakdown}
    \label{subsec: numerical fdtd simulation}
    
    \begin{figure}[htbp]
        \centering
        \includegraphics[width=1\textwidth]{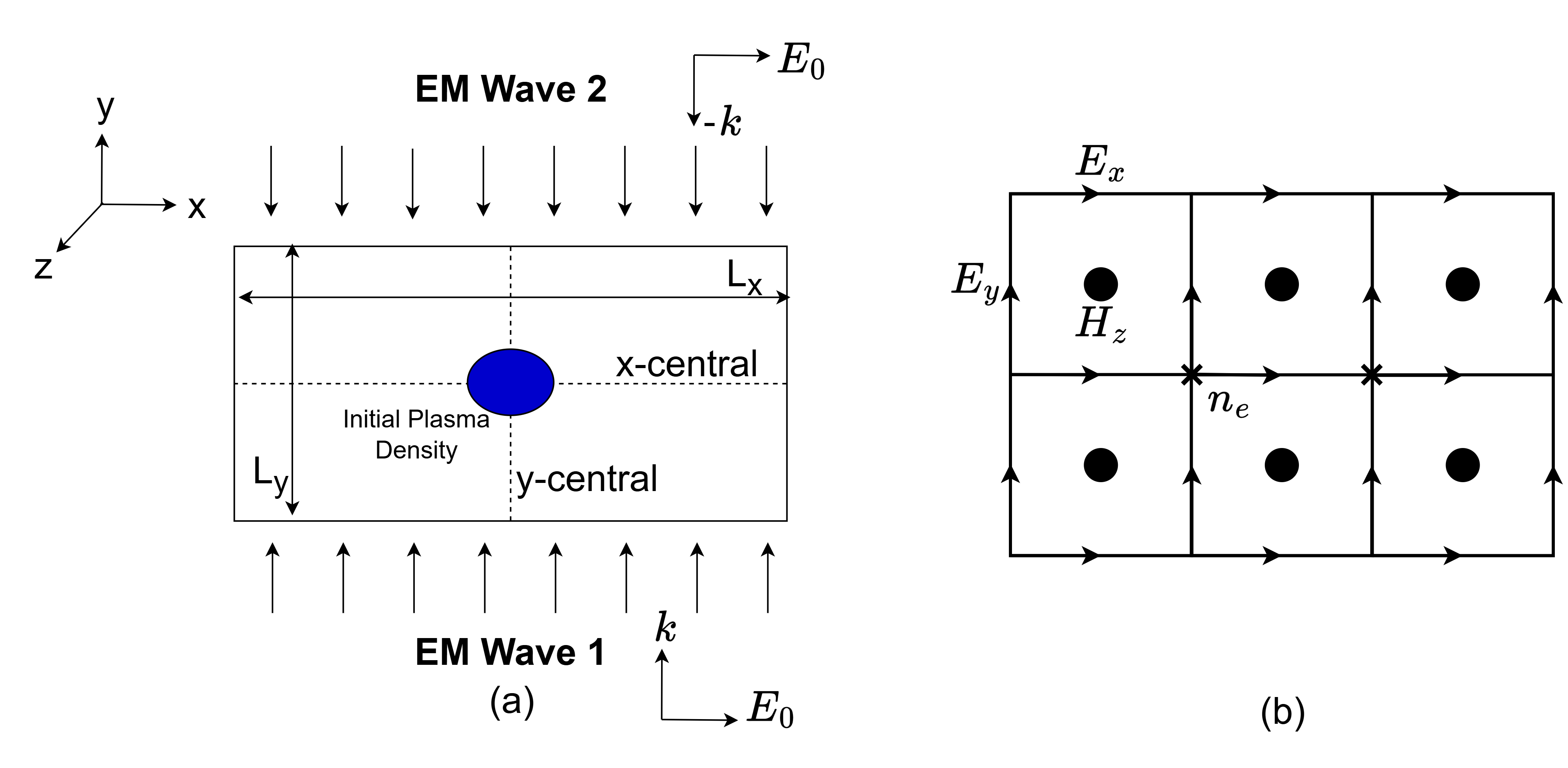}
        \caption{
        (a) Schematic of the computational domain for the HPM breakdown simulation. The domain spans \(L_x = 1\lambda\) and \(L_y = 0.5\lambda\), with an initial Gaussian plasma centrally located at \((L_x/2, L_y/2)\). Dotted lines, x-central and y-central are two lines passing through the center of the computational domain. Two identical linearly polarized plane waves (110 GHz), with electric field oriented along the \(x\)-axis, are incident from the top and bottom (\(\pm y\)-direction) with opposite wave vectors. (b) Schematic of the 2D FDTD grid illustrating the locations of the electric field (E), magnetic field (H), and electron density ($n_e$). $E_{rms}$ and ($n_e$) are calculated at the same locations on the grid.
        }
        \label{fig:domain_diagram}
    \end{figure}

    \begin{figure}[htbp]
        \centering
        \includegraphics[width=1\textwidth]{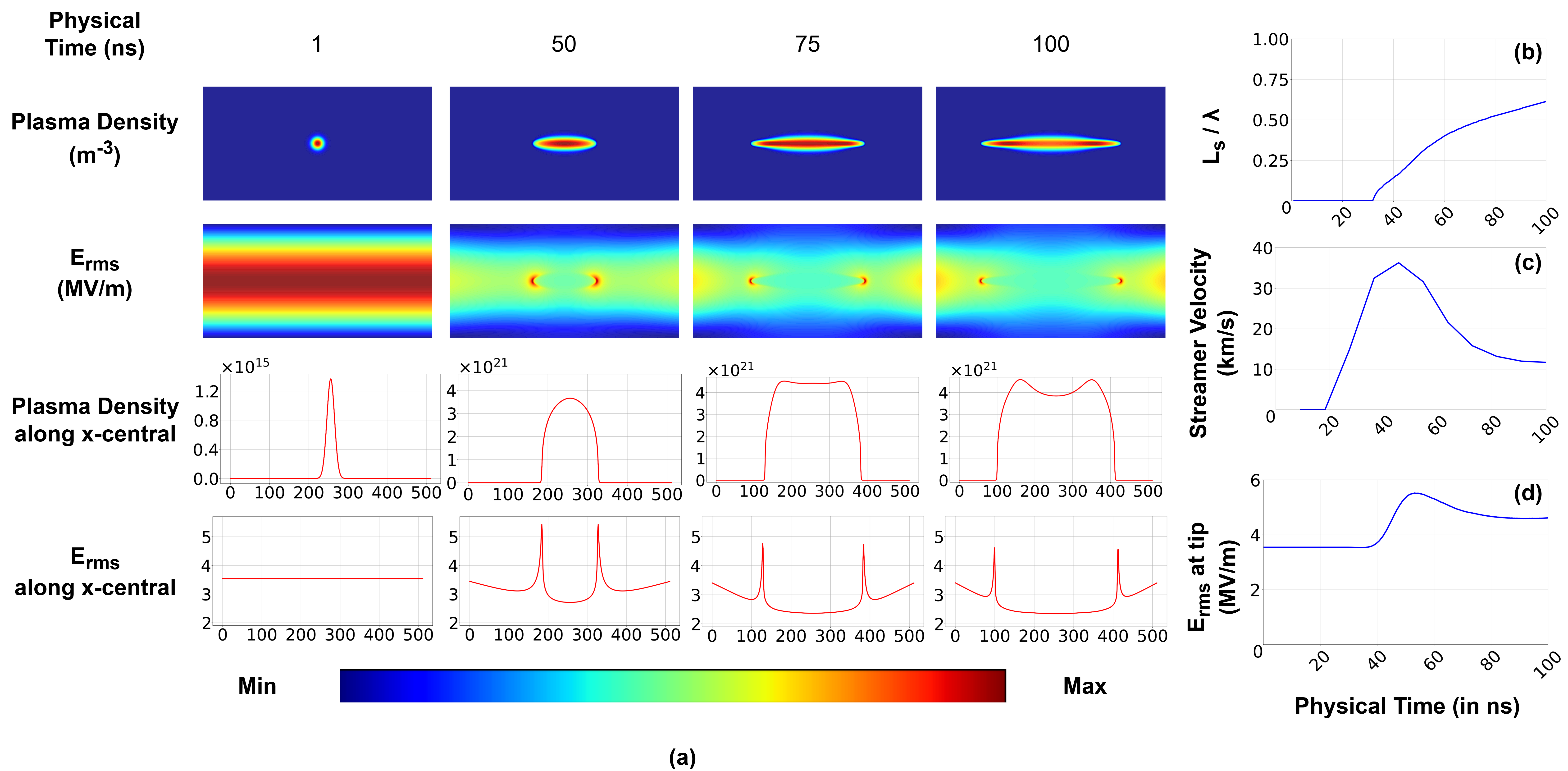}
        \caption{
            (a) Spatiotemporal evolution of plasma density (top row) and scattered electric field (second row) at different physical times, for incident electric field of \(E_0 = 2.5\,\mathrm{MV/m}\) as described in Figure~\ref{fig:domain_diagram}. The initial Gaussian plasma density evolves into an elongated streamer aligned with the incident field direction. The third and fourth rows show corresponding midline cross-sectional profiles of plasma density and electric field, respectively.  
            (b) Temporal evolution of the normalized streamer length \(L_s / \lambda\), where \(L_s\) denotes the streamer length, highlighting its growth rate and saturation over time.  
            (c) Temporal variation of the streamer growth rate as a function of time.
            (d) Temporal evolution of the scattered \(E_{\mathrm{rms}}\) at the plasma streamer tip as a function of physical time.
        }
        \label{fig:domain_sample}
    \end{figure}

    The simulations, using a conventional fluid-based approach, presented in this study are two-dimensional, with Maxwell's equations solved in a rectangular 2D computational domain using the finite difference time domain (FDTD) method based on the scattered field formulation~\cite{Kunz-1993}. In this formulation, the electric and magnetic fields are decomposed into incident and scattered components, such that the total fields are expressed as $E_{t}=E_{i}+E_{s}$ and $H_{t}=H_{i}+H_{s}$, where subscript $i,s,t$ denote incident, scattered and total fields respectively. The scattered-field approach~\cite{Kunz-1993} generates the incident wave analytically at each field vector location. The total field at any time is obtained by adding the numerically computed scattered field to the analytically calculated incident field. The FDTD method is an explicit, second-order accurate, time-domain scheme employing centered finite differences on a Cartesian grid (Figure~\ref{fig:domain_diagram} b), which provides the spatio-temporal evolution of the E and H fields~\cite{Yee-1966,gedney-2011}. The discrete E and H fields are staggered in both time and space, meaning they are shifted by half a time step and half a spatial step. The FDTD time step $\Delta t$ and grid spacing $\Delta s$ are selected to resolve the smallest relevant physical length scale, i.e. the minimum skin depth that determines electromagnetic field penetration into the plasma ~\cite{BChauduryIEEEplasma2010}. In our simulations, we use the Maxwell time step $\Delta t=.5 \Delta s/c $, which satisfies the CFL stability condition, where $c$ is the speed of light.
    The mean electron velocity equation is discretized using the direct integration scheme~\cite{BChauduryIEEEplasma2010}, as shown below:
    \begin{equation}
    \frac{v^{n+1}-v^{n}}{\Delta t}+\nu_{m}\frac{v^{n+1}+v^{n}}{2}=-\frac{e}{m}\frac{E_{t}^{n+1}+E_{t}^{n}}{2}
    \end{equation} 
    Using the above with leapfrog approximations of Maxwell's equation we get  
    \begin{equation}\label{eq:dscr1}
    E_{s}^{n+1}=E_{s}^{n}\frac{1-\beta}{1+\beta}+\frac{en_{e}\Delta t}{2\varepsilon_{0}}\frac{1+\alpha}{1+\beta}v^{n}
    -\frac{\beta}{1+\beta}(E_{i}^{n+1}+E_{i}^{n})+\frac{\Delta t}{(1+\beta)\varepsilon_{0}}\nabla\times H_{s}  \hspace{.5cm}
    \end{equation}
    \begin{equation}
    v^{n+1}=\alpha v^{n}-\frac{e\Delta t}{2m\gamma}(E_{t}^{n+1}+E_{t}^{n})\\
    \end{equation}
    \begin{equation}
    \alpha=\frac{1-a}{1+a}, \hspace{.4cm}\beta=\frac{\omega_{p}^{2}\Delta t^{2}}{4\gamma},
    \hspace{.4cm}\gamma=1+a,\hspace{.4cm} a=\frac{\nu_{m}\Delta t}{2}
    \vspace{.4cm}
    \end{equation}

    here, the temporal index '$^n$' corresponds to time t=$n\Delta t$. The magnetic field equations are advanced in time in conjunction with those for the electric field and velocity, resulting in an alternating update of the electric and magnetic field components at each grid point.
    
    In our 2D simulations, the electric field (E) lies in the plane of the computational domain, while the magnetic field (H) is oriented perpendicular to it. We consider non-zero components $E_{x}$, $E_{y}$, $H_{z}$ arranged such that 
    E is positioned along the grid edges and H on the grid faces (Fig.\ref{fig:domain_diagram}). The grid dimensions in the X and Y directions are denoted by $N_x$ $\&$ $N_y$, with electron velocity evaluated along the E-field edges and electron density positioned at cell centers, marked by ‘x’ in Fig.\ref{fig:domain_diagram} b.
    The FDTD update follows a leapfrog-in-time scheme~\cite{BChauduryIEEEplasma2010}, E is advanced from $t_n=(n-1/2)\Delta t$ to $t_n=(n+1/2)\Delta t$ using the curl of H at $t_n=n \Delta t$, after which H is advanced from $t_n=n \Delta t$  to $t_n=(n+1) \Delta t$ using the curl of E at $t_n=(n+1/2) \Delta t$. At the computational boundaries, where neighboring field components fall outside the domain, Mur’s absorbing boundary conditions are applied to compute the electric field values~\cite{Mur-1981}.

    The density equation (Plasma Solver in Figure \ref{fig:HPM_Breakdown}) is solved using a simple explicit scheme, expressed as follows:

    \begin{equation}
    \label{con:disc}
    \begin{aligned}
    n_{e,(i,j)}^{\,n+1}
    &= \frac{1}{\,1+\Delta t_{F}\bigl(\nu_{a}+r_{ei}\,n_{e,(i,j)}^{\,n}\bigr)} \Bigg[
          n_{e,(i,j)}^{\,n}\bigl(1+\Delta t_{F}\nu_{i}\bigr) \\
    &\qquad\qquad\qquad
        + \frac{D_{eff}\,\Delta t_{F}}{\Delta s_{F}^{2}}
          \bigl(n_{e,(i+1,j)}^{\,n}+n_{e,(i-1,j)}^{\,n}+n_{e,(i,j+1)}^{\,n}+n_{e,(i,j-1)}^{\,n}
          -4\,n_{e,(i,j)}^{\,n}\bigr)
    \Bigg]
    \end{aligned}
    \end{equation}
    
    where, $\Delta t_F$ and $\Delta s_F$ represents fluid time-step and grid-spacing respectively. 
    
    When determining $\Delta s_F$ and solving the continuity equation, the plasma density gradient or characteristic scale length must be taken into account. The fluid time step for the continuity equation is then obtained from the CFL condition, 
    $\Delta t_F < (\Delta s_F )^2/(2D_{max})$, where $D_{max}$ denotes the maximum effective diffusion.
    Since the characteristic timescale of plasma evolution is much longer than the EM wave period ($\approx$100 GHz), it is not necessary to solve the Maxwell and fluid equations using identical time steps. Maxwell’s equations are solved over one EM wave cycle (represented by $T_M$) using the plasma density evaluated at the start of the cycle. The plasma density is then updated for the subsequent cycle using transport coefficients that depend on the RMS field ($E_{rms}$) obtained from the previous cycle. Consequently, the density equation is integrated over a duration equal to $T_M$ using the time-averaged RMS field value $E_{rms}$ for that cycle. The RMS field is computed at the locations marked with $'\times'$ in Fig.~\ref{fig:domain_diagram}, which correspond to the electron density evaluation points. The spatial resolution for both the Maxwell and continuity equations is determined by the characteristic scale lengths of the electric field and plasma density, which in this case are of the same order. Therefore, the same grid spacing ($\Delta s_F=\Delta s$) is used for both solvers and $\Delta t_F= n\Delta t=\Delta T$, where n denoted the number of EM-solver iterations within a single EM wave period. As reported in Refs.~\cite{BChauduryIEEEplasma2010,BCHAUDHURY2018,ghosh2023efficient}, plasma density and electric field gradients can be extremely steep, requiring a very fine spatial resolution, on the order of $\lambda /500$, to capture them accurately. This fine-grid requirement for both the plasma density and Maxwell’s equations greatly increases computational cost, with accurate simulations of practical problem sizes potentially taking several months on a serial processor. \\
    Figure~\ref{fig:domain_diagram}(a) shows the simulation domain, discretized with \(512 \times 256\) grid points to satisfy the minimum spatial resolution requirement of \(N_\lambda = 500\) grid points per wavelength. The computational domain dimensions are normalized to the EM wavelength, \(L_x = 1\lambda\) and \(L_y = 0.5\lambda\). For the representative simulation presented in this section, two identical linearly polarized EM waves at 110 GHz (wavelength $\lambda$=2.7mm) with amplitude \(E_0 = 2.5 \times 10^6\, \mathrm{V/m}\) are injected symmetrically from the top and bottom boundaries, generating a standing wave field pattern aligned along the \(x\)-axis. A 2D Gaussian profiled plasma density is initialized at the domain center (predefined breakdown spot) with peak electron density \(n_0 = 10^{15}\,\mathrm{m}^{-3}\) and standard deviation \(\sigma = 5 \times 10^{-5}\, \mathrm{m}^{-3}\) forming the initial condition for breakdown onset in air at atmospheric pressure (760 Torr).
    For breakdown to occur, the electric field must exceed the critical density value at which ionization balances attachment and diffusion losses. In this setup, the maximum RMS field of the standing wave is $\approx 3.5 \times 10^{6}V/m$ (= $\sqrt{2}E_{0}$) , resulting from the superposition of two incoming waves each with amplitude $E_{0}=2.5 \times 10^{6}V/m$. This is well above the critical field of approximately 2.5 MV/m for air at atmospheric pressure in the simulations. As mentioned earlier, the EM and plasma solvers are iteratively coupled with separate time steps to update the rapidly evolving fields using a fine time step \(\Delta t\), while the plasma solver operates on a coarser step \(\Delta T = n \Delta t\). This multirate approach enables accurate resolution of both fast EM wave dynamics and slower plasma evolution. At each plasma update step, the effective electric field \(E_{\mathrm{eff}}\) is computed in the EM-solver and fed into the plasma solver, which returns the updated electron density for use in EM-solver. The simulation is terminated once the plasma streamer spans 80\% of the computational domain length in the x-direction.
    
    The plasma density distributions at different stages of the evolution of the microwave streamer for the representative case described above have been presented in the first row in Figure~\ref{fig:domain_diagram} (a). When the plasma density is low and does not significantly perturb the EM field, its spatial distribution remains Gaussian. As ionization increases plasma density toward the cutoff-density ($8.6 \times 10^{20} m{-3}$ for 110 GHz in our case), the applied field begins to be modified by the plasma. Polarization effects then enhance the electric field at the poles of the plasmoid in the field direction, leading to increased ionization in these regions. This causes the plasmoid to rapidly elongate along the field direction, forming a microwave streamer. The plasmoid tip elongates at a velocity that increases with the applied field, with an average streamer elongation velocity of approximately few tens of km/s (tip velocity on the order of $10^{6} cm/s$), as shown in the figure~\ref{fig:domain_diagram} (c). The streamer length is computed by counting the number of grid points along the $x$-central line with electron density exceeding $10^{20}\,\mathrm{m}^{-3}$, and converting this to a physical length, while the streamer growth rate is obtained by taking the temporal gradient of the streamer length and converting the result to $\mathrm{km/s}$. Figure ~\ref{fig:domain_diagram} (a) shows the 2D distributions of the RMS field and plasma density, along with the field and density profiles along the central x-axis, at four different time instants under the same conditions as Figure 2 (a). The density profile at the streamer front is extremely sharp, with a characteristic gradient length on the order of a micrometer. Overall, the simulations reveal the formation of a plasmoid that elongates along the incident field direction through a combined diffusion–ionization mechanism, eventually developing into a plasma filament or microwave streamer. Due to the high plasma density within the streamer channel, the electric field at the streamer tip is substantially enhanced, driving rapid elongation in the incident field direction. The enhanced field amplitude at the streamer head oscillates over time during elongation ~\ref{fig:domain_diagram} (c), resulting in oscillations in streamer velocity. These oscillations are linked to resonant effects.
    Despite its accuracy, the standard FDTD solver is computationally expensive. For a suare computational domain of \(1\lambda \times 1\lambda\) with \(N_\lambda = 512\), a single 2D simulation requires approximately five days of runtime on a modern desktop using serial execution. The computational complexity scales as \(\mathcal{O}(N_f)\), where \(N_f = c_k^2 N_\lambda^2\) is the total number of grid points and \(c_k\) accounts for domain scaling. Thus, scaling to a \(10\lambda \times 10\lambda\) domain would require close to 500 days, motivating the development of accelerated DL-based hybrid computational approaches without compromising fidelity \cite{ghosh2023efficient}.

\section{Hybrid FNO-Based Modeling Framework}
\label{sec:hybrid_fno_framework}

\begin{figure}[htbp]
    \centering
    \includegraphics[width=\textwidth, height=0.2\textheight, keepaspectratio]{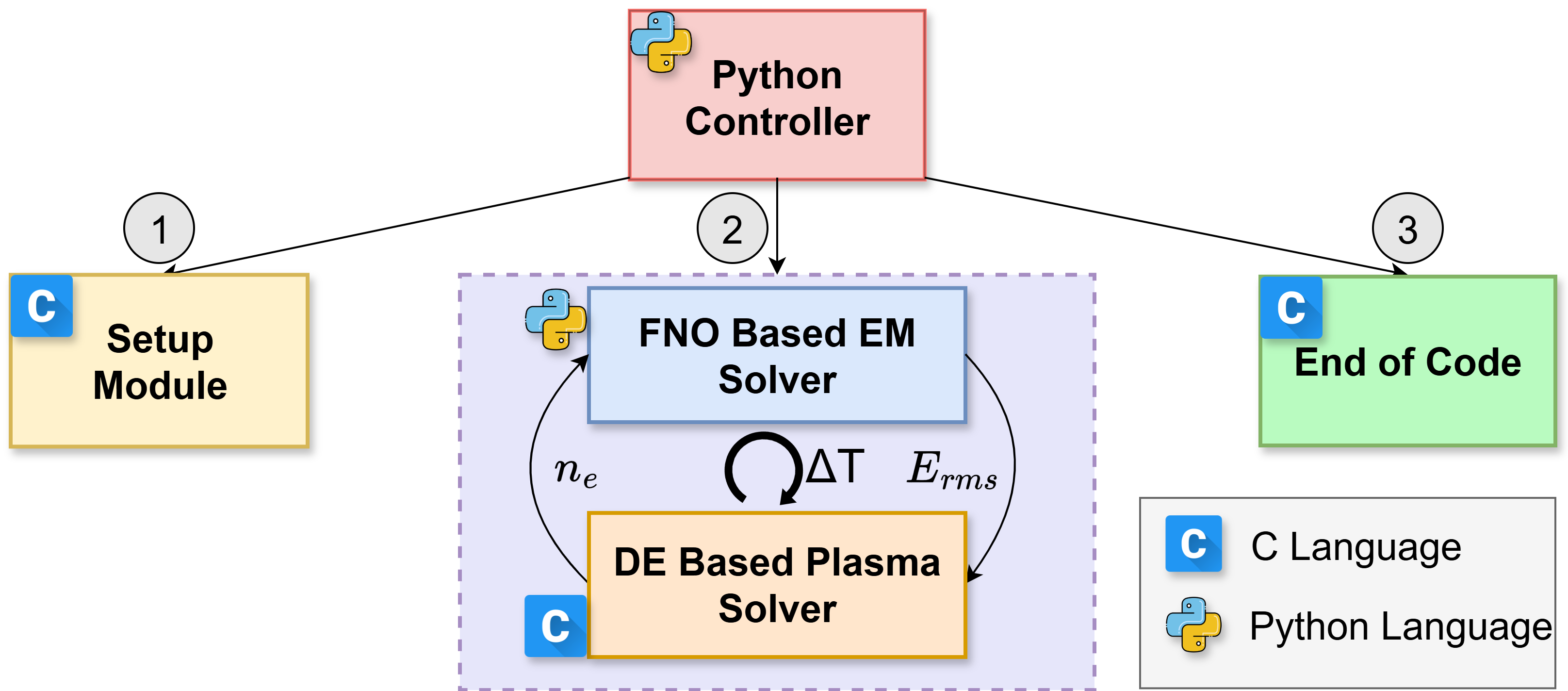}
    \caption{ Overview of the modular architecture of the deep learning-based hybrid simulation framework. The workflow shows Python used for high-level control and integration of performance optimized C-based components for numerical routines.}
    \label{fig:hybrid_simulation}
\end{figure}

Our proposed hybrid simulation framework for computational investigation of HPM breakdown seamlessly combines data-driven and physics-based solvers as shown in figure ~\ref{fig:hybrid_simulation}. Within this framework, the FNO~\cite{li2020fourier} is employed as a DL surrogate model that replaces the conventional EM solver (Block A in Figure ~\ref{fig:HPM_Breakdown}).To account for the EM-plasma interaction, the FNO surrogate is trained to predict the scattered electric field ($E_{rms}$) using the plasma density and incident field as inputs. In this formulation, the FNO is applied as an image-to-image reconstruction model, taking as input a two-channel image comprising the plasma density and incident electric field distributions, and predicting the corresponding scattered electric field image with high accuracy. This replacement offers a significant reduction in computational overhead while maintaining the physical accuracy necessary to simulate plasma dynamics in HPM breakdown scenarios.\\
Originally, the HPM breakdown solver was developed as a monolithic C codebase to exploit the computational efficiency of low-level numerical routines ~\cite{BCHAUDHURY2018}. However, integrating a DL-based surrogate necessitated restructuring the solver into a modular, controllable framework. Python was selected as the top-level controller due to its rich ML libraries like pytorch-tensorflow, native interlanguage interfacing support, and efficient data handling capabilities. To support this transition, the FDTD-based C solver was decomposed into standalone subroutines that could be dynamically invoked from Python, enabling flexible integration and decoupling of simulation components. The simulation workflow begins with a C-implemented \textit{Setup Module}, invoked from Python to initialize the simulation environment. This module defines the computational grid, spatial discretization parameter \(N_\lambda\), EM wave parameters (frequency and amplitude), ambient pressure and collision frequency. It also constructs the initial plasma density profile using a 2D Gaussian distribution, characterized by user-defined peak density and spatial variances. Once initialized, the simulation enters an iterative loop where the FNO surrogate and the C-based plasma solver operate in a tightly coupled fashion. At each iteration \(i\), the FNO module predicts \( E_{\mathrm{rms}}(i) \) and passes it to the plasma solver. The plasma solver updates the electron density \( n_e(i) \) by solving the discretized plasma continuity equations. The updated density is then returned to the FNO module, completing one simulation cycle. This loop continues until a predefined stopping condition is met. The termination criterion remains consistent with the full-physics FDTD model described in Section~\ref{sec: HPM Breakdown Simulation} and is reached once the plasma streamer spans 80\% of the domain width, corresponding to a physical length of \(0.8\lambda\). This finalization routine is implemented in C and called from Python, completing the simulation lifecycle. During execution, global parameters, such as model weights, simulation configuration, and loop counters are managed by the Python controller, while local state variables remain encapsulated within their respective C subroutines. The hybrid nature of this setup requires careful attention to data sharing between Python and C, including memory alignment, data type compatibility, and pointer referencing. These technical details are discussed in the next subsection.

\subsection{Python–C Interfacing for Hybrid Execution}
\label{subsec: py_c_interfacing}

\begin{figure*}[htbp]
    \centering
    \includegraphics[width=\textwidth]{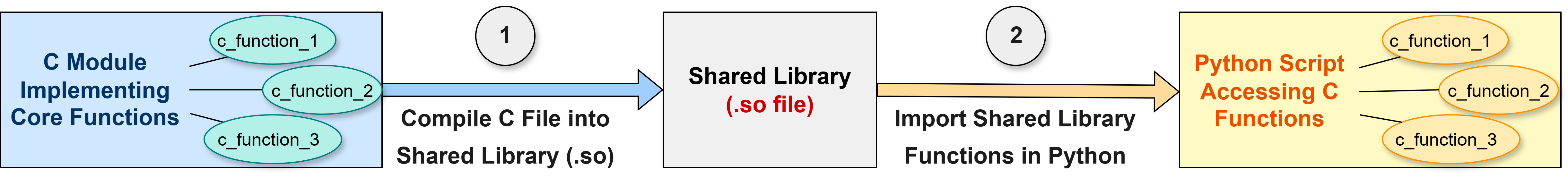}
    \vspace{-3mm}
    \caption{ Pipeline for integrating the Python-based DL surrogate EM solver with the C-based plasma solver using shared library linking. Step~1 compiles selected C functions into a platform-compatible shared object (\texttt{.so}) file. Step~2 loads the shared object into Python for execution. }
    \vspace{-3mm}
    \label{fig:py_c_pipeline}
\end{figure*}

To enable efficient cross-language integration, the proposed framework utilizes dynamic linking of shared object libraries, allowing performance-optimized C routines to operate within a Python-controlled environment. As illustrated in Figure~\ref{fig:py_c_pipeline}, the interface is implemented in two primary stages. First, selected C functions such as the plasma density solver are compiled into platform-compatible shared object (\texttt{.so}) files using GCC with the \texttt{-shared} and \texttt{-fPIC} flags. This generates dynamically loadable, position-independent binaries. In the second stage, the shared objects are loaded into Python via the \texttt{ctypes} library, which provides low-level access to C functions and memory structures, supporting pointer manipulation and type-safe argument passing.

A representative example is the function \texttt{elec\_dens()}, which updates the plasma density based on a 2D input electric field array and an integer iteration index. The function is written in C, compiled into a shared library, and invoked from Python as shown below:

\vspace{0.5em}
\begin{minipage}[t]{\linewidth}
\begin{lstlisting}[language=Python, caption={Example Python code demonstrating the invocation of the C-based \texttt{elec\_dens()} plasma density update routine through \texttt{ctypes} in the proposed hybrid simulation framework.}
]
import subprocess
import ctypes
import os
import numpy as np
from numpy.ctypeslib import ndpointer

den_py = np.zeros((256, 512))   # Initialize density matrix
itr = 0                         # Iteration index
path = os.getcwd()

# Step 1: Compile the C file into a shared library (.so)
subprocess.call(['gcc', '-shared', '-o', 'fdtd.so', '-fPIC', 'final_c_script.c', '-lm'])

# Step 2: Load the shared library into Python
fdtd = ctypes.CDLL(os.path.join(path, 'fdtd.so'))

# Step 3: Define the C function signature
elec_dens = fdtd.elec_dens
elec_dens.argtypes = [ndpointer(dtype=np.float64, ndim=2, flags="C"), ctypes.c_int]

# Step 4: Call the C function
elec_dens(den_py, itr)
\end{lstlisting}
\end{minipage}
\vspace{1em}

This approach offers three main advantages: (i) it preserves the performance benefits of low-level numerical routines already implemented in C, (ii) it enables seamless integration with Python-based ML components such as the FNO model, and (iii) it ensures minimal overhead when sharing data between the two languages, as NumPy arrays use contiguous memory buffers compatible with C.

\section{Fourier Neural Operator (FNO) Model}
\label{sec: FNO model preparation}

Unlike traditional convolutional networks, FNOs operate in the spectral domain and can learn complex, nonlocal solution operators with improved efficiency and generalization. Their ability to capture global spatial dependencies makes them particularly well-suited for problems governed by partial differential equations (PDEs). In the context of our problem, the FNO model is trained to predict the scattered electric field distribution generated from incident EM wave on a plasma, based on input fields consisting of plasma density and the corresponding incident wave pattern. The model architecture, training pipeline, and dataset design are described in detail in the following subsections.

    \subsection{Data Generation}
    \label{subsec: data_generation}
    
   The data required for model training and testing has been generated using the FDTD-based HPM breakdown simulation described in subsection~\ref{subsec: numerical fdtd simulation}. The training-validation dataset includes HPM simulations with incident electric fields of 2.5, 2.6, 2.7, 2.8, 2.9, and 3.0~MV/m, all at a fixed frequency of 110~GHz. To distinguish between different electric field amplitudes, the model input is constructed as a two-channel field - the plasma density profile as the first channel and the EM wave pattern of the corresponding incident electric field as the second. For testing the model and the simulation pipeline, unseen electric field values of 2.55, 2.65, 2.75, 2.85 and 2.95~MV/m have been employed. These test cases span the entire range of the considered electric field amplitudes (2.5–3.0~MV/m). Although these five electric field values are used exclusively for testing, the benchmarking results confirm that the proposed hybrid model generalizes well to any electric field within the specified range. Table~\ref{tab:data_distribution} lists the number of samples collected in the dataset for each incident electric field values, along with the corresponding physical simulation time.
    
    \begin{table}[htbp]
    \centering
    \caption{Number of data samples (pairs of 2D plasma density profiles and $E_{rms}$) generated from simulations using the conventional EM–plasma fluid model for various incident EM field amplitudes ($E_0$)}
    \label{tab:data_distribution}
    \resizebox{\linewidth}{!}{%
    \begin{tabular}{lcc}
    \toprule
    \textbf{Electric Field (MV/m)} & \textbf{Number of Samples} & \textbf{Physical Simulation Time (ns)} \\
    \midrule
    2.50 & 11,343 & 103.12 \\
    2.55 & 10,330 & 93.91 \\
    2.60 & 9,419 & 85.63 \\
    2.65 & 8,613 & 78.30 \\
    2.70 & 7,900 & 71.82 \\
    2.75 & 7,251 & 65.92 \\
    2.80 & 6,656 & 60.51 \\
    2.85 & 6,123 & 55.66 \\
    2.90 & 5,644 & 51.31 \\
    2.95 & 5,213 & 47.39 \\
    3.00 & 4,824 & 43.85 \\
    \bottomrule
    \end{tabular}
    }
    \end{table}
    
    The final dataset, prepared for training and testing, consists of input data with two channels, i.e. images of plasma density profile and the incident electric field, and the corresponding output data in the form of scattered field images using RMS values. Min–max normalization have been applied separately to each data type. For $E_{\text{rms}}$ and incident electric field matrices, the global minimum and maximum values are computed over the entire training and testing sets. Similarly, the plasma density profiles are normalized using global min–max values computed specifically from the density data across both training and testing sets. After normalization, all matrices are rescaled to the range of 0 to 255 to facilitate transformation of the matrix prediction task into an image generation task, leveraging the fact that convolutional layers perform better on image-like inputs.

\begin{figure}[htbp]
        \centering
        \includegraphics[width=\textwidth]{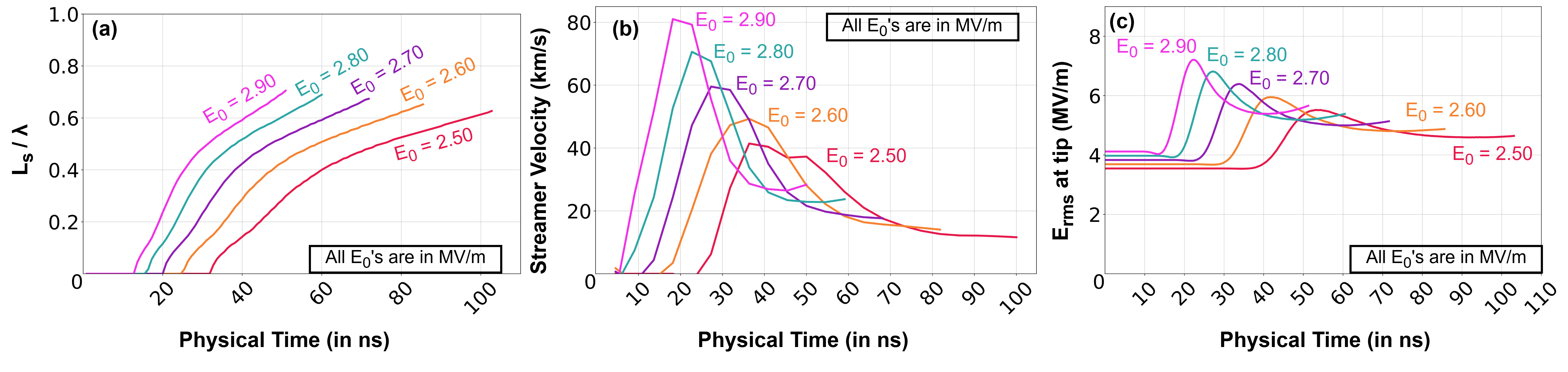}
        \caption{
            Temporal evolution of key plasma streamer parameters for different $E_0$ values. 
            (a) normalized streamer length $L_s / \lambda$,  
            (b) streamer growth rate over time, and  
            (c) scattered $E_{\mathrm{rms}}$ at the plasma tip.  
            All $E_0$ values are in MV/m.
        }

        \label{fig:fdtd_dynamics}
    \end{figure}

    Figure~\ref{fig:fdtd_dynamics} illustrates the temporal evolution of three representative plasma breakdown parameters for different incident electric field amplitudes, $E_0$. Subplot~(a) shows that the normalized streamer length $L_s / \lambda$ increases more rapidly with time as $E_0$ increases, reflecting the enhanced streamer propagation under stronger fields. Subplot~(b) presents the corresponding streamer growth rate, where higher $E_0$ values lead to earlier and more pronounced peaks, indicative of an accelerated breakdown onset. Subplot~(c) shows the scattered $E_{\mathrm{rms}}$ at the plasma tip, which exhibits both higher peak magnitudes and earlier rise times with increasing $E_0$. Collectively, these results demonstrate that larger incident field amplitudes intensify breakdown dynamics, yielding a broader range of nonlinear behaviors and providing a diverse, physically rich dataset for training the FNO-based surrogate.
    
    Following the normalization and rescaling process, the test dataset comprises a total of 37,530 input–output pairs with input shapes of (37,530, 256, 512, 2) and output shapes of (37,530, 256, 512, 1). The training dataset includes 36,864 input–output pairs, and the validation dataset contains 8,922 such pairs. The training and validation datasets are employed to train the FNO model, while the test dataset is exclusively used to evaluate both the model’s standalone performance and the performance of the proposed hybrid model discussed in Section~\ref{sec:hybrid_fno_framework}.

    \subsection{FNO Model Architecture}
    \label{subsec: model architecture}
    
\begin{figure}[htbp]
        \centering
        \includegraphics[width=1\textwidth]{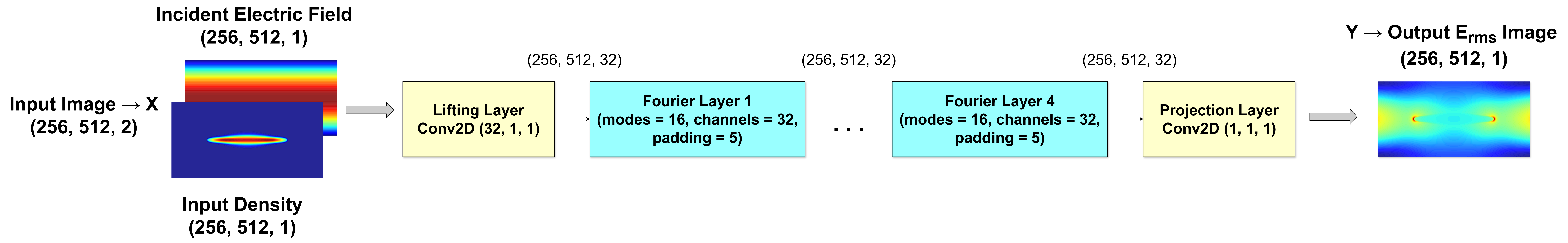}
        \caption{Overview of the FNO model pipeline. The input consists of a two-channel image comprising the normalized plasma density and incident electric field. A lifting layer maps the input to a 32-channel latent space. This is followed by four Fourier layers that perform spectral transformations and local corrections. A projection layer converts the latent representation to a single-channel output corresponding to the predicted scattered electric field.}
        \label{fig:fno_architecture}
    \end{figure}

    The structure of the FNO model used in our work is illustrated in Figure~\ref{fig:fno_architecture}. The input is a two-channel field consisting of the normalized plasma density and the incident electric field, both rescaled to the range [0, 255] prior to training. A lifting layer implemented using a $1 \times 1$ convolution expands the input into a 32-channel latent space. This lifted representation is then passed through four Fourier layers, each configured with 16 frequency modes and zero-padding of size five. The final output is reduced to a single-channel scattered electric field prediction via a projection layer, also using a $1 \times 1$ convolution.
    
    \subsubsection{Fourier Layer Design and Operation}
    
\begin{figure}[htbp]
        \centering
        \includegraphics[width=1\textwidth]{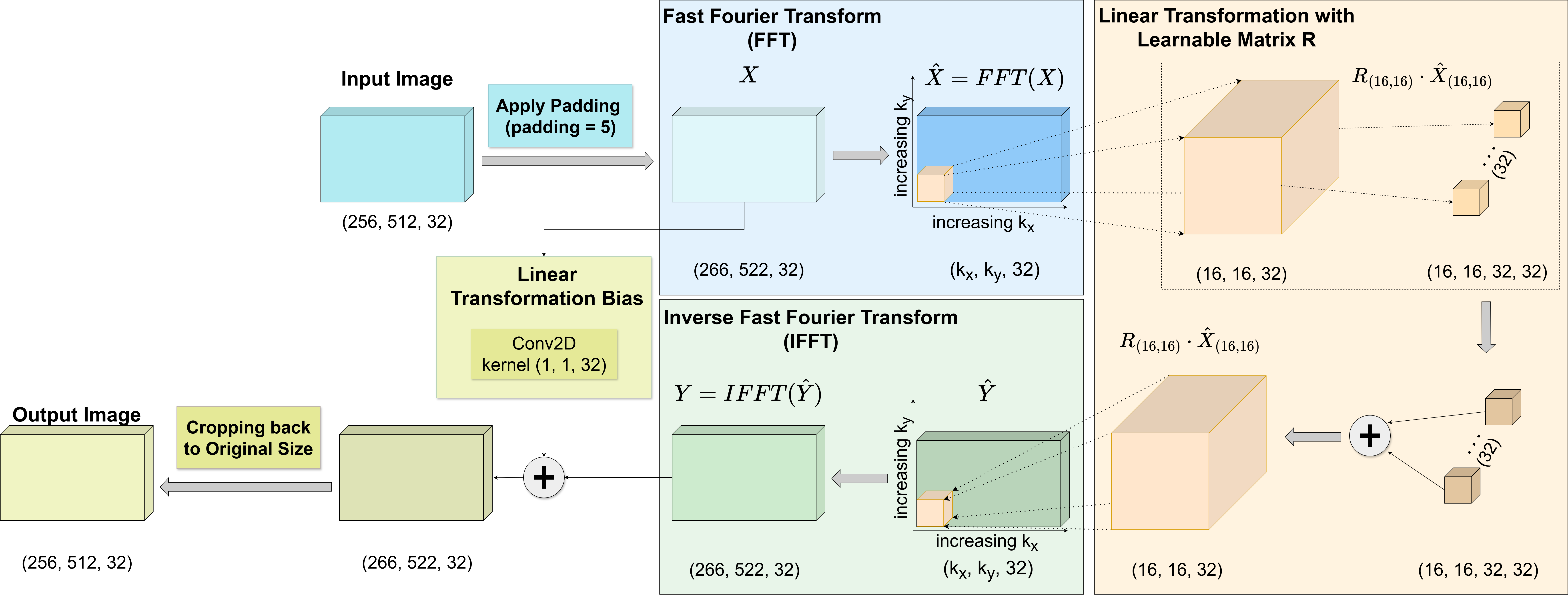}
        \caption{
        Working of a single Fourier layer. The input tensor (plasma density and related feature maps) is padded and transformed using 2D FFT. Low-frequency modes are linearly transformed using a learnable complex-valued matrix to capture global plasma–field coupling. After applying the inverse FFT, in linear transformation bias a local pointwise convolution is added to account for localized spatial plasma features. The result is cropped back to the original spatial dimensions.
        }
        \label{fig:fno_layer}
    \end{figure}

    In our configuration, each Fourier layer receives latent feature maps of size H = 256, W = 512, and C = 32, which are derived from the physical inputs of plasma density \(n_e(x,y)\) and the incident electric field. These feature maps capture the coupled spatial dependencies that determine the scattered field \(E_{\mathrm{rms}}(x,y)\). The sequence of operations is illustrated in Figure~\ref{fig:fno_layer}.
    
    Let \(\mathbf{X} \in \mathbb{R}^{H \times W \times C}\) denote the input tensor to a Fourier layer. First, zero-padding of size five is applied along the spatial dimensions to reduce boundary information loss. The tensor is then transformed into the spectral domain via a two-dimensional Fast Fourier Transform (FFT):
    
    \begin{equation}
    \hat{\mathbf{X}}(k_x, k_y) = \mathcal{F}[\mathbf{X}(x, y)] = \sum_{x=0}^{H-1} \sum_{y=0}^{W-1} \mathbf{X}(x, y) e^{-2\pi i \left( \frac{k_x x}{H} + \frac{k_y y}{W} \right)}.
    \end{equation}
    here, \((k_x, k_y)\) represent discrete frequency indices in the \(x\) and \(y\) directions. In our configuration, only the lowest \(m = 16\) Fourier modes in each direction are retained, \(\hat{\mathbf{X}}_{|k_x|<m, |k_y|<m}\), as these capture the dominant long-range interactions relevant to plasma dynamics while reducing computational cost.
    
    A learnable complex-valued weight matrix \(\mathbf{R} \in \mathbb{C}^{m \times m \times C \times C}\) is then applied
    \begin{equation}
    \hat{\mathbf{Y}}(k_x, k_y) = \mathbf{R}(k_x, k_y) \cdot \hat{\mathbf{X}}(k_x, k_y),
    \end{equation}
    for \(|k_x|<m\) and \(|k_y|<m\), leaving higher modes unchanged.
    
    The modified spectral tensor \(\hat{\mathbf{Y}}\) is then zero-padded back to the full frequency resolution and transformed into the spatial domain using the inverse FFT.
    
    \begin{equation}
    \mathbf{Y}(x, y) = \mathcal{F}^{-1}[\hat{\mathbf{Y}}(k_x, k_y)] = \frac{1}{HW} \sum_{k_x=0}^{H-1} \sum_{k_y=0}^{W-1} \hat{\mathbf{Y}}(k_x, k_y) e^{2\pi i \left( \frac{k_x x}{H} + \frac{k_y y}{W} \right)}.    
    \end{equation}
    
    To reintroduce spatial nonlinearity, which is critical for modeling localized streamer tips and plasma boundary effects, a pointwise convolution (\(1 \times 1\) kernel) is applied as a bias term and added to \(\mathbf{Y}\). Finally, the output is cropped to the original spatial dimensions (H, W) before being passed to the next Fourier layer.
    By performing global spectral transformations, each Fourier layer captures how the large-scale distribution of \(n_e\) modulates the scattered field \(E_{\mathrm{rms}}\), while the linear transformation bias accounts for fine-scale streamer features and non-periodic boundary effects. This integration of frequency-domain learning with localized corrections allows the model to resolve both global EM–plasma interactions and linear transformation bias learning localized nonuniform spatial plasma features, both of which are essential to accurately model HPM breakdown physics.
    
    \subsection{Training Protocol and Implementation}
    \label{subsec: training loop}
    
    The FNO model is implemented using the open-source \texttt{physicsnemo} library developed by NVIDIA, which provides modular support for physics-informed neural operator architectures. Training is performed using PyTorch on NVIDIA RTX 6000 Ada Generation GPU. The model is trained using a batch size of 32 with the Adam optimizer and an initial learning rate of $10^{-3}$. Training is run for 300 epochs with early stopping triggered by stagnation in validation loss.
    
\begin{figure}[htbp]
        \centering
        \includegraphics[width=0.9\textwidth]{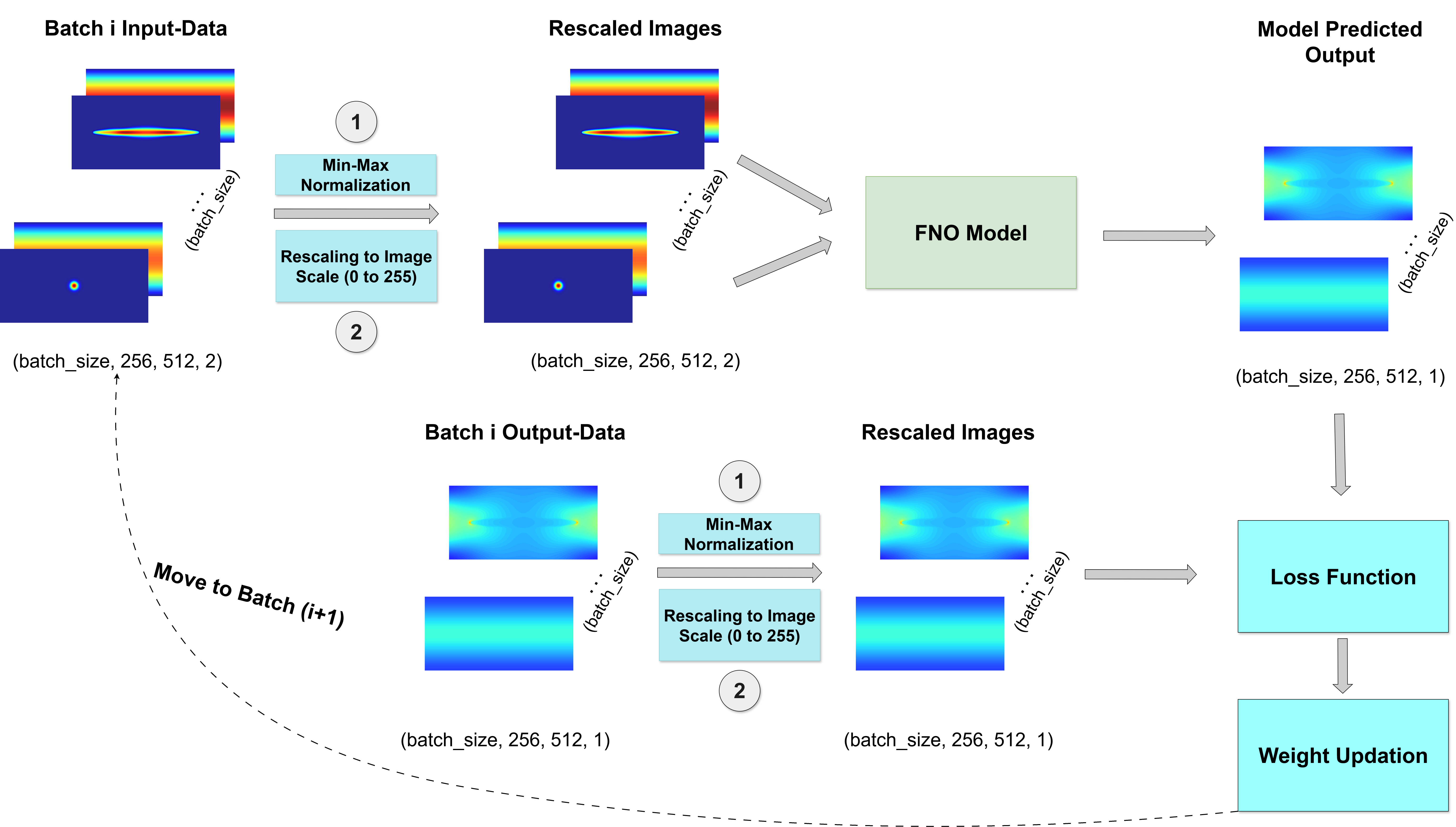}
        \caption{Training loop for the FNO model. Input and ground-truth output fields are normalized and rescaled. The predicted field is compared with the target using the MSE loss. Model weights are updated via backpropagation. Performance is evaluated using SSIM and APE to monitor physical accuracy and perceptual similarity.}
        \label{fig:fno_training}
    \end{figure}

    Mean squared error (MSE) is used as the primary loss function, ensuring accurate pixel-wise reconstruction of the scattered electric field. Additionally, structural similarity index (SSIM) and average percent error (APE) are computed throughout training to assess both perceptual quality and relative scaling accuracy. These supplementary metrics offer insights into how well the model preserves large-scale field structure and physical correctness. The full training and evaluation loop is summarized in Figure~\ref{fig:fno_training}. At each epoch, mini-batches of input-output pairs are processed, and gradients are backpropagated to update weights. Validation is conducted on a held-out subset, as mentioned in subsection~\ref{subsec: data_generation}.
    
    This FNO-based surrogate provides a compact and efficient means of capturing electromagnetic response in distinct plasma streamer profiles. Its ability to generalize to unseen incident field profiles, while maintaining agreement with FDTD-based fluid solvers, makes it a suitable tool for using it in HPM breakdown instead of computationally expensive DE based EM solver.

\section{Results and Validation}
\label{sec: results_validation}

Evaluation and validation have been performed in two stages. In the first stage, we perform quantitative and qualitative evaluation of the performance of FNO model. This step provides confidence in replacing the classical DE-based EM solver with the FNO-based solver. Since the instantaneous scattered electric field prediction must be highly accurate to carry forward the entire simulation, even a small discontinuity or mismatch in prediction of $E_{rms}$ could lead to uncontrolled plasma streamer growth over long durations. To verify robustness, a second stage of simulation-based evaluation is conducted, comparing the results of the FNO based hybrid approach with those from conventional FDTD based plasma fluid model. This stage wise performance evaluations are discussed in detail in the following subsections. To ensure generalization, all performance evaluation simulations are run on unseen test data, with incident fields of 2.55, 2.65, 2.75, 2.85, and 2.95~MV/m, as discussed in subsection~\ref{subsec: data_generation}.

\subsection{Performance Analysis of FNO Model}
\label{subsec:fno_performance}

\begin{figure}[htbp]
    \centering
    \includegraphics[width=1\textwidth]{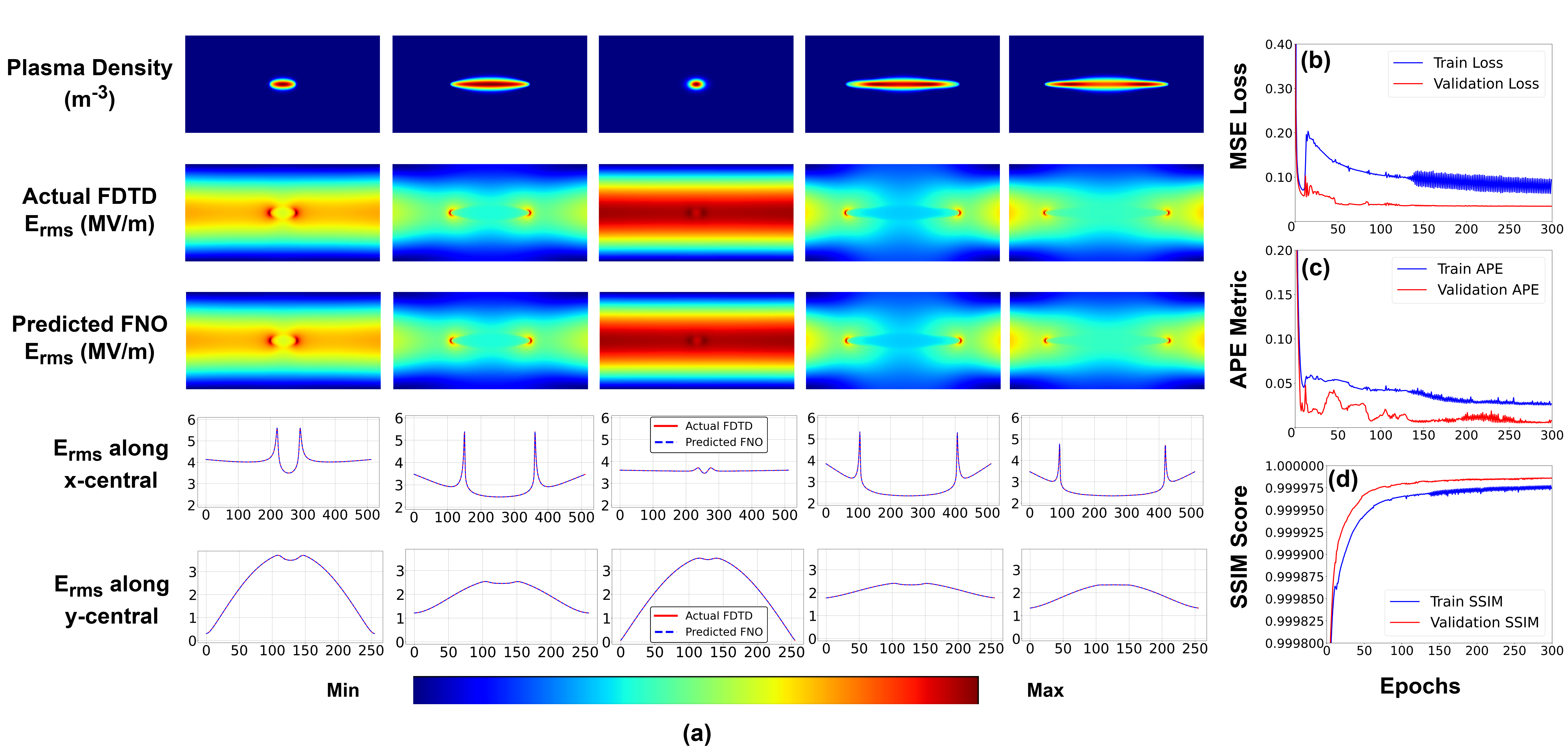}
    \caption{Evaluation of FNO model performance. (a) FNO model predicted vs. actual FDTD based $E_{\text{rms}}$ fields with 2D and 1D comparisons, and (b)--(d) training–validation curves for MSE loss, APE, and SSIM metrics.}
    \label{fig:fno_model_performance}
\end{figure}

We performed two types of assessment to evaluate this model performance, (i) a metrics-based assessment using MSE, SSIM, and APE, and (ii) a qualitative confirmation where 2D and 1D plots are compared with ground truth results. These evaluations are crucial, as even a very small error can lead to large-scale deviations over the course of a full simulation because ionization which is dependent on the local effective field is a non-linear phenomenon during plasma evolution. In addition to the MSE loss function, APE and SSIM metrics are used for pixel-based and structural-based comparisons, respectively. APE compares predicted results pixel-by-pixel with the ground truth, while SSIM performs a structure-wise comparison between ground truth and predicted results. For a test dataset consisting of 37,530 input–output image pairs, the FNO model achieved an average MSE of 0.0394, an average APE of 0.12, and an average SSIM of 0.9999, showing excellent agreement with the actual results.\\
Figure~\ref{fig:fno_model_performance}(a) presents a qualitative comparison between FNO predicted and actual FDTD generated $E_{\text{rms}}$ data for different electric field and density profile inputs. The high similarity between the results obtained from two methods is evident in both the 2D maps and the 1D profiles along the x-central. Figure~\ref{fig:fno_model_performance}(b)--(d) show how training and validation metrics evolve over 300 epochs, MSE and APE consistently decrease, while SSIM increases, indicating stable and accurate learning.

\subsection{Performance Analysis of Hybrid FNO based Fluid Model}
\label{subsec:hybrid_performance}

\begin{figure}[htbp]
    \centering
    \includegraphics[width=1\textwidth]{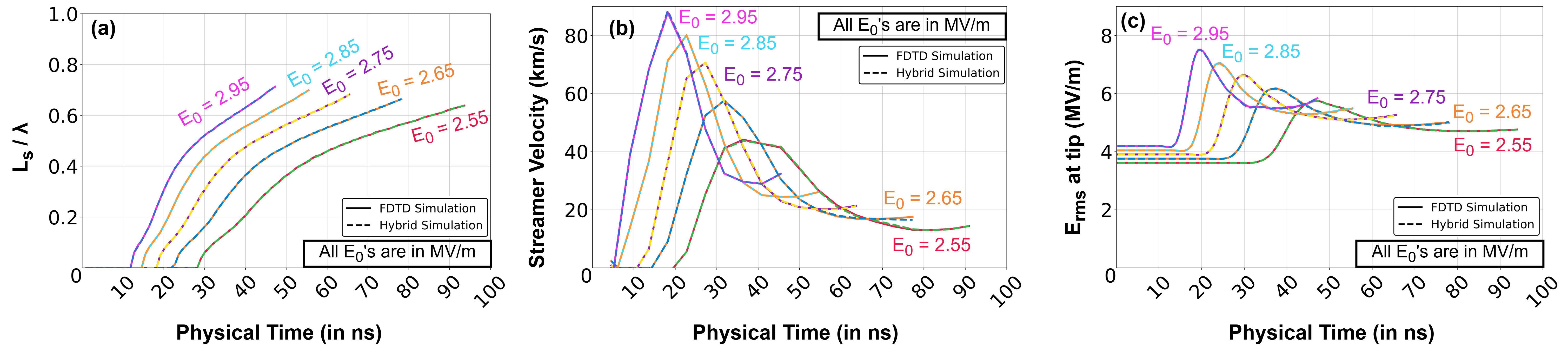}
    \caption{Comparison of temporal evolution of key plasma streamer parameters for different $E_0$ values between the FNO-based hybrid model and the FDTD-based plasma fluid model.  
        (a) normalized streamer length $L_s / \lambda$,  
        (b) streamer growth rate over time, and  
        (c) scattered $E_{\mathrm{rms}}$ at the plasma streamer tip.  
        All $E_0$ values are in MV/m.}
    \label{fig:filament_growth}
\end{figure}

Following the validation of scattered electric field predictions in subsection~\ref{subsec:fno_performance}, the second stage of evaluation examines the temporal evolution performance of the proposed FNO-based hybrid model. The goal is to verify that the hybrid solver can accurately reproduce plasma streamer evolution due to ionization-diffusion mechanism over extended simulation durations, without introducing numerical instabilities or deviating from the established FDTD-based plasma fluid model. To this end, both the hybrid and FDTD solvers were executed on identical, previously unseen test cases with incident electric field amplitudes of $E_0 = 2.55$, $2.65$, $2.75$, $2.85$, and $2.95$~MV/m. Figure~\ref{fig:filament_growth} compares key parameters relevant to HPM breakdown physics, namely,   
(a) normalized streamer length $L_s / \lambda$,  
(b) streamer growth rate as a function of time, and  
(c) scattered $E_{\mathrm{rms}}$ at the streamer tip.  
Across all cases, the hybrid solver predictions exhibit an almost perfect overlap with those from the FDTD-based plasma fluid model. This high level of agreement capturing both amplitude and temporal evolution indicates that the hybrid model has successfully learned and reproduced the underlying physics of the HPM breakdown process.

\begin{figure}[htbp]
    \centering
    \includegraphics[width=1\textwidth]{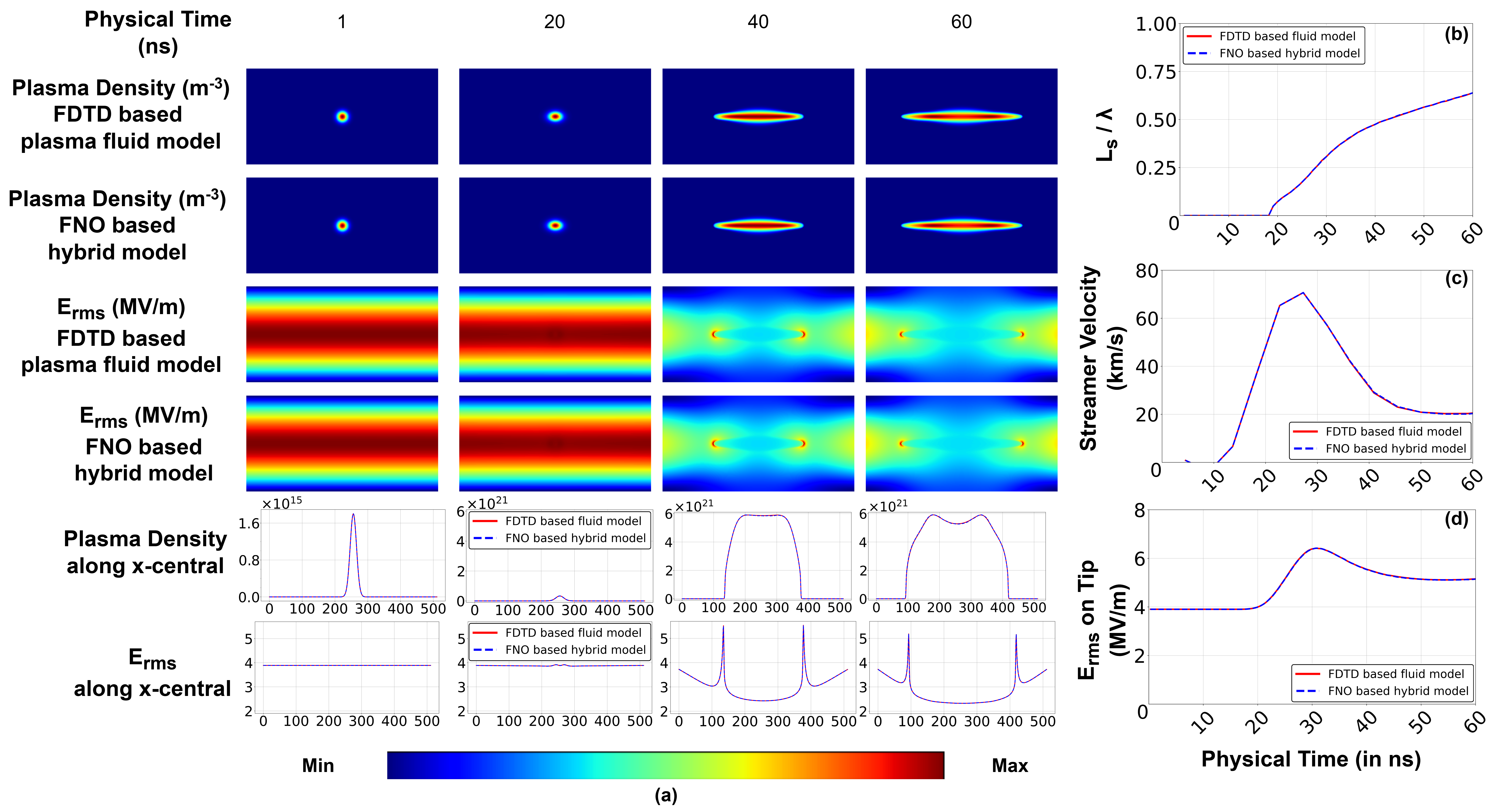}
    \caption{Comparison of plasma density and scattered $E_{\mathrm{rms}}$ fields for $E_0 = 2.75$~MV/m between the full FDTD based plasma fluid simulation and the proposed FNO-based hybrid fluid simulation. Both 2D maps and 1D $x$-central cuts exhibit excellent agreement between DE based conventional simulation and data-driven hybrid simulation approaches. (b-d) Evolution of plasma streamer length, streamer velocity and $E_{rms}$ on plasma streamer tip shows excellent match between the two approaches throughout the complete simulation.}
\label{fig:hybrid_one_electric_field}
\end{figure}

A more detailed spatial comparison is presented in Figure~\ref{fig:hybrid_one_electric_field}, which shows the plasma streamer evolution for $E_0 = 2.75$~MV/m at multiple physical time steps. The 2D spatial maps of plasma density and scattered $E_{\mathrm{rms}}$ fields from the hybrid solver are in excellent agreement with those from the FDTD based fluid simulation, both in terms of morphology and amplitude. Additionally, 1D central $x$-axis cut profiles further confirm that the hybrid model reproduces the FDTD based fluid model results with high fidelity in both amplitude and phase.

In addition to accuracy, computational efficiency is a major strength of the hybrid framework. Table~\ref{tab:speedup_with_table} summarizes the runtime comparison between the two approaches. For all $E_0$ values tested, the hybrid solver achieves speed-up factors ranging from $57.10\times$ to $62.77\times$, reducing multi-hour FDTD simulations to under $40$ minutes without sacrificing accuracy. As discussed in subsection~\ref{subsec: numerical fdtd simulation}, scaling this problem to a domain size of $10\lambda \times 10\lambda$ would require approximately $500$ days of computation using the FDTD-based plasma fluid model~\cite{ghosh2023efficient}. With the average speed-up obtained from hybrid solver, the same large-scale simulation could be completed in few days. This represents a drastic reduction in computational cost, effectively transforming a simulation that would take over 1.5 years into one that can be completed in less than a week.

\begin{table}[htbp]
    \centering
    \caption{Runtime performance comparison between FDTD and hybrid simulations for different $E_0$ values.}
    \label{tab:speedup_with_table}
    \resizebox{\linewidth}{!}{%
    \begin{tabular}{lccc}
    \toprule
    \textbf{Electric Field (MV/m)} & \textbf{FDTD Simulation Time (min)} & \textbf{Hybrid Simulation Time (min)} & \textbf{Speed-Up} \\
    \midrule
    2.55 & 2131.67 & 37.34 & 57.10$\times$ \\
    2.65 & 1637.82 & 26.69 & 61.28$\times$ \\
    2.75 & 1395.00 & 22.20 & 62.77$\times$ \\
    2.85 & 1184.34 & 19.89 & 59.44$\times$ \\
    2.95 &  991.28 & 16.98 & 58.32$\times$ \\
    \bottomrule
    \end{tabular}
    }
\end{table}

Overall, these results demonstrate that the proposed FNO-based hybrid model not only preserves the physical accuracy of the full-wave FDTD solver but also delivers substantial computational gains. This combination of fidelity and efficiency positions the hybrid solver as a powerful tool for large-scale and long-duration plasma streamer simulations in high-power microwave breakdown studies.

\section{Conclusion}\label{sec: Conclusion} A hybrid data-driven computational framework for multiscale multiphysics simulation of HPM breakdown at high pressures involving high frequency microwaves has been presented. Compared to differential equation (DE) based numerical approach, in this framework, FNO–based deep learning (DL) surrogate replaces the computationally intensive FDTD based EM solver for Maxwell's equation within a widely used EM–plasma fluid model. By retaining the established plasma continuity equation solver for electron density evolution, the approach maintains the physical accuracy while achieving substantial computational acceleration. To the best of our knowledge, this is the first reported application of such hybridization in the context of plasma fluid modeling and simulation of HPM breakdown. The hybrid model has been trained and validated using datasets generated from high-fidelity FDTD–plasma fluid simulations of microwave streamer formation over a range of incident electric field amplitudes, with independent test cases chosen to evaluate generalization capability. The FNO surrogate demonstrated excellent agreement with reference solutions for scattered electric field predictions due to EM-plasma interactions, achieving average SSIM values approaching unity, minimal APE, and low MSE. When integrated into the coupled EM–plasma fluid framework, the FNO augmented hybrid data-driven modeling approach reproduced streamer shape, growth rate, and temporal evolution due to diffusion-ionization mechanism with near-perfect overlap compared to the conventional full-physics DE based EM-plasma fluid solver, for all tested cases.

A key outcome of this work is the significant performance improvement, the hybrid approach achieved speedup factors between $55\times$ to $60\times$, reducing multi-day simulations to under few hours without loss of accuracy. Such acceleration enables simulations of larger computational domains (e.g., $10\lambda \times 10\lambda$) and longer physical timescales that would otherwise be computationally prohibitive with conventional methods. The efficiency gains are particularly valuable for multiscale, multiphysics plasma simulation problems where EM field updates dominate runtime. From an implementation standpoint, the modular Python–C integration strategy developed here allows seamless coupling of high-performance compiled solvers (in C, C++, or Fortran) with modern machine learning frameworks. This facilitates the adoption of the hybrid paradigm in existing legacy plasma simulation codes with minimal restructuring, extending its applicability beyond HPM breakdown to a wide range of plasma-assisted processes, gas discharges, and other EM–plasma interaction scenarios. In summary, this work establishes a generalizable and scalable hybrid modeling framework that combines the accuracy of physics-based solvers with the computational efficiency of machine learning surrogates for advancing plasma science and engineering.

\section{Acknowledgment}
We acknowledge BSES Rajdhani and BSES Yamuna for CSR grants to carry out the work at the SELC, DAU, Gandhinagar, India. The authors acknowledge Kaushal Patel for his contributions towards the development of the Python–C interfacing for Hybrid Execution. The authors acknowledge Libin Varghese for fruitful discussion on this project.

\bibliographystyle{abbrv}
\bibliography{article}

\end{document}